# Ordered vs disordered: Correlation lengths of 2D Potts models at $\beta_t$

Wolfhard Janke and Stefan Kappler[a][*]

[a]Institut für Physik, Johannes Gutenberg-Universität Mainz, 55099 Mainz, Germany

We performed Monte Carlo simulations of two-dimensional $q$-state Potts models with $q = 10, 15$, and 20 and measured the spin-spin correlation function at the first-order transition point $\beta_t$ in the disordered and ordered phase. Our results for the correlation length $\xi_d(\beta_t)$ in the disordered phase are compatible with an analytic formula. Estimates of the correlation length $\xi_o(\beta_t)$ in the ordered phase yield strong numerical evidence that $R \equiv \xi_o(\beta_t)/\xi_d(\beta_t) = 1$.

## 1. INTRODUCTION

Correlation lengths or inverse masses are among the most important quantities to characterize the properties of a statistical system. Unfortunately there are only very few models for which the correlation length $\xi$ is exactly known and can thus serve as a testing ground for the employed numerical techniques. The best known example is the two-dimensional (2D) Ising model where $\xi$ is exactly known at all temperatures. But the situation is already much more involved for 2D $q$-state Potts models with a partition function

$$Z = \sum_{\{s_i\}} e^{-\beta E}; \; E = -\sum_{\langle ij \rangle} \delta_{s_i s_j}; \; s_i = 1,\ldots,q, \quad (1)$$

where $i = (i_x, i_y)$ denote the lattice sites of a square lattice of size $V = L_x \times L_y$, $\langle ij \rangle$ are nearest-neighbor pairs, and $\delta_{s_i s_j}$ is the Kronecker delta symbol. Here only the correlation length $\xi_d(\beta_t)$ in the disordered phase could be derived analytically [1], where $\beta_t = \ln(1 + \sqrt{q})$ is the first-order transition point of this model for $q \geq 5$. For the correlation length $\xi_o(\beta_t)$ in the ordered phase no analytical results are available. One goal of the present investigations was to test the conjecture that $R \equiv \xi_o(\beta_t)/\xi_d(\beta_t) = 1/2$, which was suggested quite heuristically by previous Monte Carlo (MC) studies [2].

[*]WJ would like to thank the DFG for a Heisenberg fellowship and SK gratefully acknowledges a fellowship by the Graduiertenkolleg "Physik und Chemie supramolekularer Systeme". Work supported by computer grants HLRZ hkf001 and NVV bvpf03.

## 2. SIMULATION

We studied the Potts model (1) in both the disordered and ordered phase and measured correlation functions at $\beta_t$ for $q = 10, 15$ and 20 on lattices of size $V = L \times L$ and $V = 2L \times L$ with $L = 150, 60$ and 40 ($\approx 14\xi_d$). To take advantage of translational invariance we used periodic boundary conditions. We carefully checked that our lattice sizes are large enough to suppress tunneling events such that, starting from a completely random or ordered configuration, the system remained a sufficiently long time in the disordered or ordered phase to perform statistically meaningful measurements. Analyses of autocorrelation times suggested to use in the disordered phase the single-cluster and in the ordered phase the heat-bath update algorithm. All error bars are estimated by means of the jack-knife technique.

To determine the correlation length $\xi$ we considered the $k_y^{(n)} = 2\pi n/L_y$ momentum projections,

$$g^{(n)}(i_x, j_x) = \frac{1}{L_y} \sum_{i_y, j_y} G(i, j) e^{ik_y^{(n)}(i_y - j_y)}, \quad (2)$$

of the spin-spin correlation function and its improved cluster estimator

$$G(i,j) \equiv \langle \delta_{s_i s_j} - \frac{1}{q} \rangle = \frac{q-1}{q} \langle \Theta(i,j) \rangle, \quad (3)$$

where $\Theta(i,j) = 1$, if $i$ and $j$ belong to the same Swendsen-Wang cluster, and $\Theta = 0$ otherwise. The projection removes the power-like prefactor in the large-distance behaviour of $G$ and thus allows fits of $g^{(n)}(x) \equiv g^{(n)}(i_x, 0)$ to the Ansatz

$$g^{(n)}(x) = a\,\mathrm{ch}(\frac{L_x/2-x}{\xi^{(n)}}) + b\,\mathrm{ch}(c\frac{L_x/2-x}{\xi^{(n)}}), \quad (4)$$

where

$$\xi^{(n)} = \xi/\sqrt{1+(2\pi n\xi/L_y)^2} \approx \xi(1+\kappa(n/L_y)^2). \quad (5)$$

In the ordered phase (4) can only be used for $n \neq 0$.

## 3. RESULTS

### 3.1. Disordered phase

A preliminary report of a first set of simulations in the disordered phase on $L \times L$ lattices was already given in Ref.[3]. In the meantime we have further increased the statistics and added another set of simulations on asymmetric $2L \times L$ lattices [4]. For both lattice geometries the first three energy moments are found in very good agreement with the exact result for the average energy and with dual transformed large $q$ expansions [4,5].

In the disordered phase we concentrated on the $k_y = 0$ projection of the correlation function (3). In a first step we fixed $\xi^{(0)} = \xi_d$ at its theoretical value (see Table 1) and optimized only the remaining three parameters in (4). For $q = 10$ the resulting fits to the $L \times L$ and $2L \times L$ data are shown in Fig. 1 as dotted and solid lines, respectively. While the lines are excellent interpolations of the data over a wide range up to $x \approx (5\ldots6)\xi_d$, we also see a clear tendency of the fits to lie systematically above the data for very large $x$. In unconstrained fits to the Ansatz (4) this is reflected by a systematic trend to underestimate $\xi_d$. By restricting the fit interval to larger $x$ values the estimates for $\xi_d$ increase, but then also the statistical errors grow rapidly [4]. Table 1 shows our results for $\xi_d(\beta_t)$ using the fit range $[x_{\min}, L/2]$ with $x_{\min} \approx 2\xi_d$. We further estimate $c \approx 1.5 - 2$, independent of $q$, which stabilizes to $c = 1.5(1)$ if $\xi_d$ is held fixed at its theoretical value.

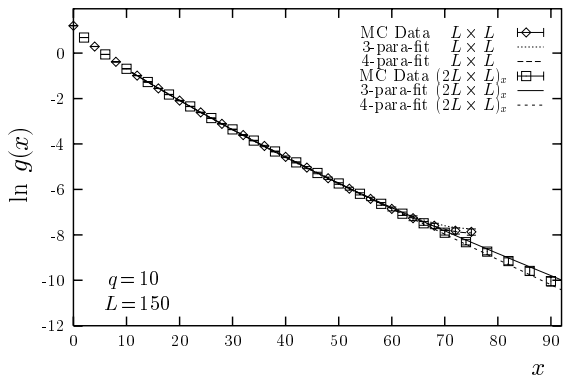

Figure 1. Semi-log plot of the projected correlation functions for $q = 10$ in the disordered phase.

Table 1
Correlation length $\xi_d(\beta_t)$ in the disordered phase.

|              | $q=10$   | $q=15$     | $q=20$   |
|--------------|----------|------------|----------|
| $L \times L$ | 9.0(5)   | 3.70(16)   | 2.24(6)  |
| $(2L \times L)_x$ | 10.2(9)  | 3.59(10)   | 2.23(5)  |
| $(2L \times L)_y$ | 9.3(7)   | 3.62(16)   | 2.33(7)  |
| exact        | 10.5595... | 4.1809... | 2.6955... |

### 3.2. Ordered phase

Also in the ordered phase we first checked that the average energy agrees with the exact result, and compared the higher moments with the large $q$ expansions of Ref.[5]. For the specific heat see Table 2. Furthermore we looked at the magnetization $m = (q\langle\max\{n_i\}\rangle/V - 1)/(q-1)$ and its cluster estimator $m' = \langle|C|_{\max}\rangle/V$, where $n_i$ denotes the number of spins of "orientation" $i = 1, \ldots, q$ and $|C|_{\max}$ is the size of the largest (spanning) cluster. Also here we find very good agreement with the exact answer [6] (for $q = 10$ and $2L \times L$, e.g., $m = m' = 0.857113(49)$, $m_{\mathrm{ex}} = 0.857106\ldots$).

To determine $\xi_o$ we followed Gupta and Irbäck [2] and studied the $k_y = 2\pi/L_y$ projection $g^{(1)}(x)$ of $G$. Since this removes constant background terms, $g^{(1)}(x)$ can be fitted with the Ansatz (4) also in the ordered phase. For $q = 10$ the qualitative behaviour of $g^{(1)}$ is illustrated in the semi-log plot of Fig. 2. The comparison with $g^{(0)}$ of the disordered phase suggests that the two correlation functions are governed by the same asymptotic decay law, i.e., that $\xi_o(\beta_t) = \xi_d(\beta_t)$. In fact, the dotted line interpolating the $g^{(1)}$ data is a constrained fit to (4) assuming that $\xi_o = \xi_d = 10.5595\ldots$. To make this statement even more convincing we have plotted in Fig. 3 the ratio $R^{\text{eff}} = \xi_o^{\text{eff}}/\xi_d^{\text{eff}}$, where $\xi^{\text{eff}} = 1/\ln[g(x)/g(x+1)]$ is the usual effective correlation length (with the correction (5) for $\xi_o$ already taken into account). The corresponding plots for $q = 15$ and $q = 20$ look similar.

Table 2
Specific heat at $\beta_t$ in the ordered phase.

|              | $q=10$     | $q=15$     | $q=20$     |
|--------------|------------|------------|------------|
| $L \times L$ | 17.95(13)  | 8.016(21)  | 5.351(15)  |
| $2L \times L$ | 17.81(10) | 8.004(19)  | 5.3612(55) |
| large $q$    | 18.1(1)    | 8.00(3)    | 5.362(5)   |

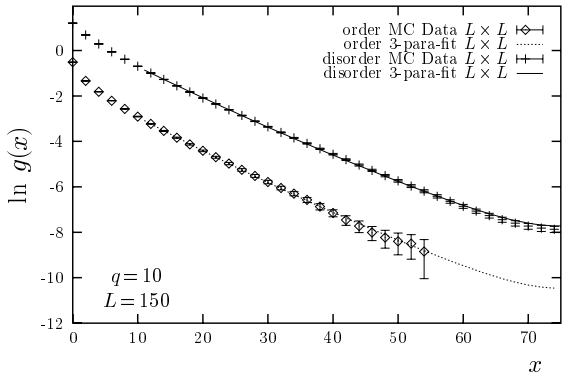

Figure 2. Semi-log plot of the projected correlation function $g^{(1)}$ for $q = 10$ in the ordered phase.

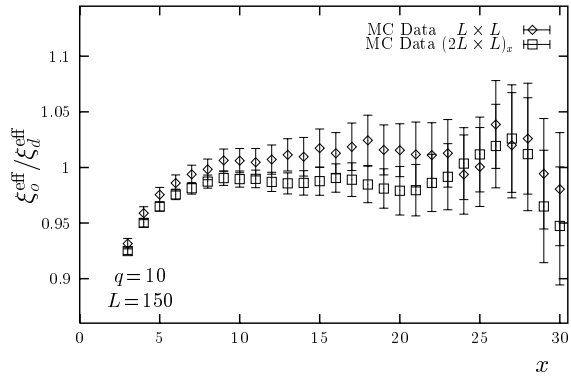

Figure 3. Ratio of effective correlation lengths in the ordered and disordered phase for $q = 10$.

## 4. DISCUSSION

The numerical results for $\xi_d(\beta_t)$ confirm the analytical expression to about $10 - 20\%$. The systematic underestimate is presumably caused by higher excitations which are neglected in the fits. Using a non-zero momentum projection in the ordered phase we believe that $\xi_o(\beta_t)$ is of about the same accuracy. By comparing the two correlation length at $\beta_t$ we obtain strong numerical evidence that $\xi_o = \xi_d$. At first sight this is in striking disagreement with a very recent exact proof [7] of the earlier conjecture $\xi_o = \xi_d/2$ for a certain definition of the ordered correlation length, $\xi_{o,1}$. For another definition, $\xi_{o,2}$, however, only the relation $\xi_{o,2} \geq \xi_{o,1}$ could be established, which would be consistent with our results if we identify the numerically determined $\xi_o$ with $\xi_{o,2}$. We are currently investigating this problem in more detail [8] by using precisely the definitions of Ref.[7] which are based on geometrical properties of Potts model clusters such as, e.g., their diameter.